# Feasibility of photoelectron sources for testing the energy scale stability of the KATRIN β-ray spectrometer

O. Dragoun [a], A. Špalek [a], J. Kašpar [a,1], J. Bonn [b], A. Kovalík [c,2],
E. W. Otten [b], D. Vénos [a,3], Ch. Weinheimer [d]

[a] *Nuclear Physics Institute, Academy of Sciences of the Czech Republic, CZ-25068 Řež near Prague, Czech Republic*
[b] *Institut für Physik, Johannes Gutenberg-Universität Mainz, D-55099 Mainz, Germany*
[c] *Laboratory of Nuclear Problems, Joint Institute for Nuclear Research, RU-141980 Dubna, Russia*
[d] *Institut für Kernphysik, Westfälische Wilhelms-Universität Münster, D-48149 Münster, Germany*

Photoabsorption of nuclear γ-rays in thin metallic convertors was examined with the aim to produce monoenergetic photoelectrons of kinetic energy around 18.6 keV and natural width of about 1 eV. Calculations were carried out for commercial photon sources of $^{241}$Am (1.1 GBq) and $^{119m}$Sn (0.5 GBq) irradiating Co and Ti convertors. Photoelectrons ejected by $^{241}$Am γ- and X-rays from Co convertors of various thickness were measured with two electrostatic spectrometers.

## 1. Introduction

Neutrinos, together with photons, are probably the most abundant elementary particles of the Universe. Still, the neutrino rest mass $m_\nu$ remains unknown. Due to great importance of this quantity for cosmology, astroparticle and particle physics [1, 2] various experimental methods have been developed [3, 4] but only upper limits for $m_\nu$ were reliably determined until now [5]. The most sensitive method for model independent determination of $m_\nu$ proved to be investigation of the uppermost part of tritium β-spectrum with endpoint energy of 18.6 keV. The present best limits of $m_\nu \leq 2.3$ eV [6] and $m_\nu \leq 2.5$ eV [7] at 95% C. L. were achieved with a new type of electron spectrometer applying a magnetic adiabatic collimation to an electrostatic filter (MAC-E filter). These instruments, combining instrumental resolution of a few eV with transmission of several tens of percent of 4π, were developed independently at the University of Mainz [8] and at the Institute of Nuclear Research at Troitsk near Moscow [9]. Detail discussion of all aspects concerning neutrino mass limit from tritium β decay can be found in recent review [10].

Since the set-ups in Mainz and Troitsk practically reached their sensitivity limits, the KATRIN (KArlsruhe TRItium Neutrino experiment) collaboration was founded in 2001 [11] with the aim to build a substantially larger MAC-E-filter than the two previous ones. Operating with a windowless gaseous tritium source (WGTS) this spectrometer should reach, after 1000 days of data taking, the sensitivity to neutrino mass of 0.2 eV [12]. One of the important conditions for reaching this ambitious goal is to guarantee a long-term stability of the spectrometer energy scale on a ±3 ppm level (e.i. ±60 meV at 18.6 keV) at least for two months – a typical individual run of the KATRIN measurements. Any larger Gaussian smearing [13] or an unrecognized shift [14] of the energy scale would result in an unacceptable systematic error of derived neutrino mass with respect to intended sensitivity.

---

[1] On leave at the Center for Experimental Nuclear Physics and Astrophysics, and Department of Physics, University of Washington, Seattle, WA 98195, USA.
[2] On leave from the NPI ASCR, Řež near Prague.
[3] Tel.: +420 212 241 677; fax: +420 220 941 130. Email address: venos@ujf.cas.cz (D. Vénos).

The instability of the KATRIN energy scale may be caused by the following reasons:
1) instability of the 18.6 kV voltage on the central retardation electrode of the MAC-E-filter
2) instability of the scanning voltage applied to the WGTS
3) changes of the electrostatic potential within WGTS
4) changes of the work function of the central retardation electrode of the MAC-E-filter.

Voltage of 18.6 keV will be measured continuously using the high-voltage divider [15] developed according to the reference divider of the Physikalisch-Technische Bundesanstalt in Braunschweig [16]. The low voltage at HV divider output as well as the scanning voltage will be measured with sufficient precision by means of commercial voltmeters. Potential distribution within WGTS will be examined by adding a small amount of radioactive $^{83m}$Kr ($T_{1/2}$ = 1.8 h) into the WGTS and measuring a possible broadening of the 17.8 keV conversion electron line populated in the isomeric decay. Long-term behavior of the spectrometer work function has to be examined in a dedicated experiment.

In accord with the KATRIN Design Report [12], there will be an additional independent monitoring spectrometer for checking stability of 18.6 keV voltage. This spectrometer will be the former Mainz MAC-E filter, reconstructed for the instrumental resolution of 1 eV. The instrument will be connected to the same high-voltage power supply as the KATRIN main spectrometer and adjusted to a sharp electron line of extremely stable energy. This monitoring line should have energy close to the tritium β-spectrum endpoint. Stable monitoring line position will indicate stability of the HV system and consequently correctness of the measured value of 18.6 keV voltage determined by the HV divider and corresponding precision voltmeter.

Development of an electron source of extremely stable energy is thus one of the KATRIN tasks, since no such energy standards are commercially available. In principle, solid state sources of internal conversion electrons, photoelectrons and Auger electrons could serve for the purpose, assuming that the natural width of the monitoring line will not exceed 2–3 eV. This, together with the instrumental resolution of a few eV, is the necessary condition to resolve in a measured electron spectrum the zero-energy-loss peak of the line. This peak corresponds to electrons that did not suffer any energy loss due to shake-up/off processes and/or inelastic scattering within the source material.

One of suitable monoenergetic electron sources is the condensed $^{83m}$Kr source [17] widely utilized in the Mainz neutrino experiment [6] and further improved [18] for extensive use in the KATRIN experiment. Under development is the $^{83}$Rb/$^{83m}$Kr solid radioactive source [19] with the half-life of 86 days, more appropriate for long-term monitoring. Both sources are emitting conversion electrons of energies 7.5 – 32 keV that are useful in testing the spectrometer resolution function and linearity. The 17.8 keV electrons of 2.7 eV natural width (the K-shell line of the 32 keV transition, the K-32 line) are well suited for testing the energy stability of the KATRIN spectrometer.

In this work, we investigate a feasibility of an energy standard utilizing photoelectrons ejected from a solid convertor by γ-rays of extremely small natural width.

## 2. Requirements and possibilities
### 2.1. Sources of electrons with stable energy

In conversion electron spectroscopy [20], similarly to the photoelectron spectroscopy, the original kinetic energy of electrons ejected from atoms in solid or gaseous phase is mainly given by the difference of the excitation energy (virtual and real photons in mentioned processes, respectively) and of the electron binding energy. The kinetic energy of electrons emitted from a solid sample and measured in the spectrometer is given by [21]:

$$E_{kin} = E_{exc} - E_{b,F} - E_{rec} - \varphi_{sp} - C, \qquad (1)$$

where $E_{exc}$ is either γ-ray or X-ray energy, $E_{b,F}$ is the electron binding energy referred to the Fermi level, $E_{rec}$ is the recoil energy (less than 0.25 eV in our case), $\varphi_{sp}$ is the spectrometer work function and $C$ is a correction for a possible charging of the sample surface.

A high stability of $E_{kin}$, necessary for monitoring of the KATRIN energy scale, requires corresponding stability of all four quantities on the right hand side of Eq. (1). Changes of γ-ray energy, investigated by the Mössbauer spectroscopy, are completely negligible in our application. The X-ray photons that could produce the 18.6 keV photoelectrons have too large natural width. Thus the chemical changes of X-ray energies, reaching sometimes tenths of eV, do not need to be considered here.

However, the electron binding energy, $E_b$, is known to depend on a physical/chemical environment of the atom emitting electron. This phenomenon is widely utilized in surface analysis by X-ray photoelectron spectroscopy [22]. It was also applied several times in high-resolution conversion-electron spectroscopy of trace amounts of radioactive $^{99m}$Tc and $^{235m}$U. For example, authors [23], applying instrumental resolution better than 1 eV, determined that the binding energies of $3d_{3/2}$ and $3d_{5/2}$ electrons of technetium in two different valence states differ by $\Delta E_b(NH_4TcO_4-TcO_2 \cdot 2H_2O) = 3.6 \pm 0.3$ eV. In subsequent studies [24], the shifts $\Delta E_b(Tc_{metal}-TcO_4) = 5.61 \pm 0.15$ eV and $\Delta E_b(Tc_{metal}-Tc_{cluster}) = 0.44 \pm 0.07$ eV were found. Notations $Tc_{metal}$ and $Tc_{cluster}$ corresponded to more than $10^{16}$ and less than $10^{14}$ technetium atoms/cm$^2$ in the samples, respectively.

Correction $C$ for a possible charging of the sample surface is an important quantity since, in the spectroscopy of zero-energy-loss electrons with energy around 20 keV, one is collecting information from the sample surface layer of thickness not exceeding 100 nm. Electrical potential of this layer may differ from that of the source bulk. This is particularly true for thin layers of radioactive substances prepared by vacuum evaporation, when trace amounts of impurities are also deposited on the backing surface (e.g. [25]).

A special feature of the $^{83}$Rb/$^{83m}$Kr source [19] is that the 17.8 keV conversion electrons are not emitted in the decay of $^{83}$Rb ($T_{1/2}$ = 86 d) but in the isomeric decay of daughter $^{83m}$Kr ($T_{1/2}$ = 1.8 h). For the $^{83}$Rb/$^{83m}$Kr source prepared by vacuum evaporation, the trace amount of $^{83}$Rb (typically $10^{13}$ atoms of a few MBq activity) will surely be present in a chemical compound – most probably oxides and hydroxides – accompanied by much larger amount of impurities. These unavoidable impurities are evaporated simultaneously with $^{83}$Rb regardless careful chromatographic procedures and strong preheating of the evaporation boat. Note that the magnitude of the chemical shifts of $E_b$ is determined by atoms nearest to that of $^{83m}$Kr emitting conversion electrons. The kind and amount of these neighbor atoms is hardly controllable. The number of $^{83}$Rb atoms in the source will decrease by a factor of two after 86 days, the time comparable with KATRIN individual measurement runs. Thus the neighbors of the $^{83m}$Kr atoms at the run beginning and end may not be identical not speaking about possible drift of $^{83m}$Kr atoms in the source surface layer prior to its isomeric decay. These effects could result in an undesirable time shift of $E_{kin}$ due to a change of $E_b$. One could expect better stability for the sources prepared by implantation of $^{83}$Rb into metallic foils.

A hydrocarbon contamination layer may slowly grow on the spectrometer electrode surface regardless good vacuum conditions which may cause a change of the spectrometer work function $\varphi_{sp}$. It must be therefore verified that possible long-term changes of $\varphi_{sp}$ do not exceed an acceptable limit or are known and predictable.

In monitoring of the spectrometer energy scale in the KATRIN experiment, not the absolute value of the electron binding energy but its time stability plays a decisive role. Still, to achieve stability of tens of meV at 18.6 keV for at least two months is a challenging task. Although a reasonable progress has been achieved in development of the conversion electron monitoring sources [18, 19] yielding requested count rates of $10^3$ s$^{-1}$ and stability at the level off 3 ppm/month it is important to examine an alternative possibility. This is why we have investigated what can be gained employing of the photoelectric effect. The physical/chemical state of a metallic convertor (with its surface cleaned by ion etching) irradiated by real photons might supply photoelectrons with more stable kinetic energy than do the conversion electrons emitted from open radioactive sources including implanted ones. In any case, this would be the source of monoenergetic electrons with other systematic errors.

**2.2. γ-ray emitters and photoelectron convertors**

Half-life of many excited nuclear states is longer than 1 ps and therefore deexciting γ-rays exhibit natural widths $\Gamma_{nucl}$ < 1 meV. Their contribution to the natural width of produced photoelectrons, $\Gamma_{phe} = \Gamma_{nucl} + \Gamma_{atom}$, is thus completely negligible. This is the main advantage of γ-rays in comparison with X-rays, the natural width of which is of the order of units or tens of eV. In order to obtain sufficient photoelectron intensity the convertor with high atomic number $Z$ is preferable. This, however, is in contradiction with the requirement for sufficiently small $\Gamma_{phe}$. As a compromise, we limited our choice to metallic convertors with $Z \leq$ 30, i.e. zinc having $\Gamma_K$ = 1.62 eV [26]. We need photoelectrons with kinetic energy $E_{kin} \approx$ 18.6 keV and since $E_{b,F}$ < 10 keV for the K-shell electrons in chosen $Z$ region, we come to the energy of exciting γ-rays, $E_{exc}$ < 30 keV. Suitable radionuclide should have half-life of at least several months and should not emit too much other γ- and X-rays that would increase background in measured photoelectron spectrum.

E. G. Kessler, Jr. [27] called our attention to the fact that γ-rays of 26.3446(2) keV energy [28] emitted in the $^{241}$Am decay would eject from the K-shell of cobalt ($E_{b,F}$ = 7.70875(80) keV [29]) photoelectrons with suitable kinetic energy of about 18.632 keV (assuming $\varphi_{sp} \approx$ 4 eV in Eq. (1)), which is just above the endpoint of 18.575 keV of the tritium β-spectrum. Another suitable property of the $^{241}$Am/Co photoelectron source would be a small natural widths of the monitoring line since the natural width of the K-shell in cobalt is of 1.28 eV [26] while the natural width of exciting γ-rays, their Doppler broadening at 300 K and recoil energy are less than 0.02 eV. However, a small decay probability of $^{241}$Am and extensive self-absorption of low-energy photons in Am (Z=95) is a limiting factor for achievable intensity of 26.3 keV γ-rays. Besides, branching ratio of the 26.3 keV photons is only 2.4 % in the $^{241}$Am decay and there are also disturbing X-rays and 59.6 keV γ-rays [30] (see section 3.2). Possible slow oxidation of an originally clean surface of the cobalt convertor should not cause principal difficulties, since the binding energy of 2p electrons in metallic cobalt differs by 2.0 and 2.1 eV from that of cobalt oxides CoO and $Co_2O_3$, respectively [31]. Photoelectrons corresponding to the metal component originate with higher energy than those corresponding to Co oxide states and can be resolved in the monitor spectrometer.

We also considered irradiation of a thin titanium foil with 23.870(8) keV γ-rays emitted in 16.5 % of the $^{119m}$Sn decays [32]. The photoelectrons knocked out from the Ti K-shell ($E_{b,F}$ = 4.964881(59) keV [29]) would likewise have suitable kinetic energy of 18.901 keV, this time 0.326 keV above the tritium β-decay endpoint. The $^{119m}$Sn half-life of 293 d is suitable for the monitoring purposes. However, there is a question of practically attainable specific activity of $^{119m}$Sn since it is usually produced in a reactor via (n, γ) reaction. Disturbing role of intensive X-rays of energy around 24 keV needs to be investigated since their photo absorption on the titanium K-shell could increase background in the neighborhood of the 18.9 keV monitoring line (see section 3.3).

In our calculations and preliminary measurements, the convertors with a diameter of 8 mm were used.

## 2.3. Commercial sources of low-energy γ-rays

We need strong γ-ray sources that could be treated as sealed radioactive sources, even in vacuum of $10^{-10}$ mbar needed for a reliable operation of the MAC-E-filter. We have chosen the $^{241}$Am source (type AMC13145, No. 1473 CW) produced by Nuclitec GmbH [33], and the $^{119m}$Sn source (type MSn9.154) made by Ritverc, GmbH [34]. Both sources are of a disc shape and have a beryllium window for smaller attenuation of low-energy photons. Properties of the two technically closed commercial sources are summarized in Table 1.

The source shown in Fig. 1 involves $^{241}$Am incorporated in a ceramic enamel placed in a welded monel capsule with beryllium exit window. The source of 1.1 GBq activity is more appropriate for our purpose than the next stronger source of 3.7 GBq also available from Nuclitec. Due to pronounced self absorption of low-energy photons in americium (see section 4.1) the stronger source would provide only ≈35 % higher intensity of useful 26.3 keV γ-rays while the intensity of disturbing 59.6 keV γ-rays would increase by a factor of 2.8. The dose equivalent rate at 1 cm distance from the 1.1 GBq $^{241}$Am source in direction of the beryllium window is 1 mSv/min [34]. This $^{241}$Am source also emits about 300 neutrons per second due to (α, n) reactions with the low-Z elements (e.g. Si, Al, O) in surrounding material.

The considered $^{119m}$Sn source had nominal activity of 555 MBq. Construction of this source is similar to that of described $^{241}$Am source. The $^{119m}$Sn activity in the chemical form of $SnO_2$ is deposited in a titanium alloy holder with exit beryllium window. Producer [34] guarantees a specific activity higher than 11 GBq/g and radioisotope purity better than 99.9 %. This specific activity (obviously the highest one that can be achieved by irradiating enriched $^{118}$Sn in a high flux reactor) is unfortunately $10^4$ times smaller than the theoretical limit determined by the $^{119m}$Sn half life of 239 d. Investigators [35] reached a very high specific activity of $^{119m}$Sn via the (α, n) reaction on $^{116}$Cd enriched to 97.07 %. However, γ-ray spectroscopy revealed impurities of $^{113}$Sn ($T_{1/2}$ = 115 d) even 6 months after the irradiation.

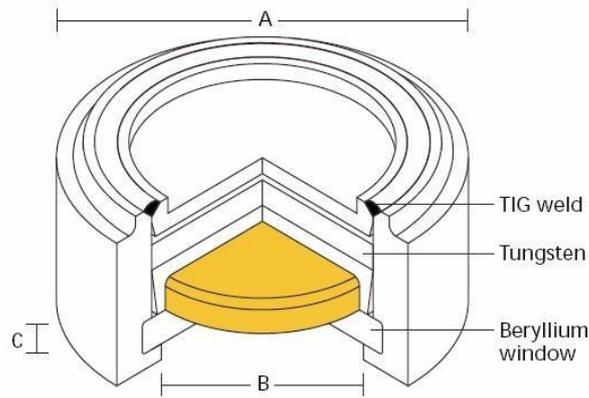

**Fig. 1.** Schematic view of the 1.1 GBq $^{241}$Am source of γ-and X-rays [34]. For the 1.1 GBq source the diameter of the capsule from a monel metal is A = 10.8 mm. Diameter of the active part of $^{241}$Am in enamel is B = 7.2 mm, thickness of the beryllium window is C = 0.95 –1.05 mm.

## 3. Calculations
### 3.1. Photons and photoelectrons emitted by the $^{241}$Am/Co and $^{119m}$Sn/Ti sources

Some of γ- and X-rays emitted during the $^{241}$Am and $^{119m}$Sn decays are absorbed in the source material and Be exit window, some of them create photoelectrons in Co or Ti convertors. These photons can also reach a detector. We are interested in counting rate of the 18.632 keV and 18.901 keV photoelectrons emitted from the Co and Ti convertor, respectively, as well as in corresponding background rates. To elucidate the role of various processes, we carry out simplified calculations, followed by several Monte Carlo simulations. The main interaction processes are shown schematically in Fig. 2 and 3.

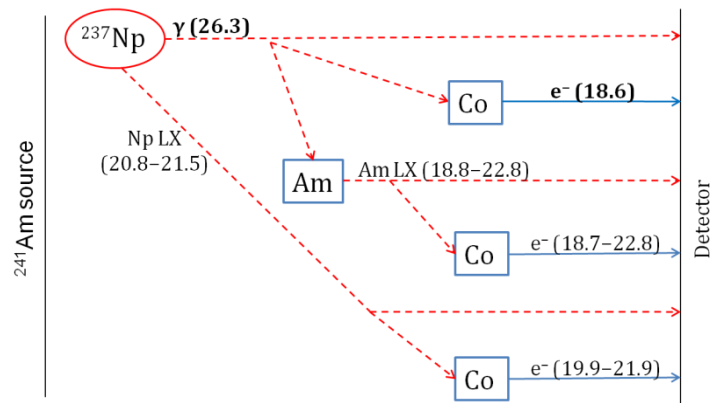

**Fig. 2.** Photons and electrons of energy around 18.6 keV emitted from the $^{241}$Am/Co photoelectron source. The α-decay of $^{241}$Am populates excited levels of the $^{237}$Np nucleus that may deexcite via emission of indicated γ-rays. Filling of the vacancies in the L-subshells of Np and Am atoms (after $^{241}$Am decay and selfabsorption of γ-rays in americium, respectively) is accompanied by emission of characteristic X-rays of the L-series. Not shown, but included into calculations are interactions with Si-like enamel, Be exit window, C backing of the Co convertor, and hydrocarbon contamination overlayer.

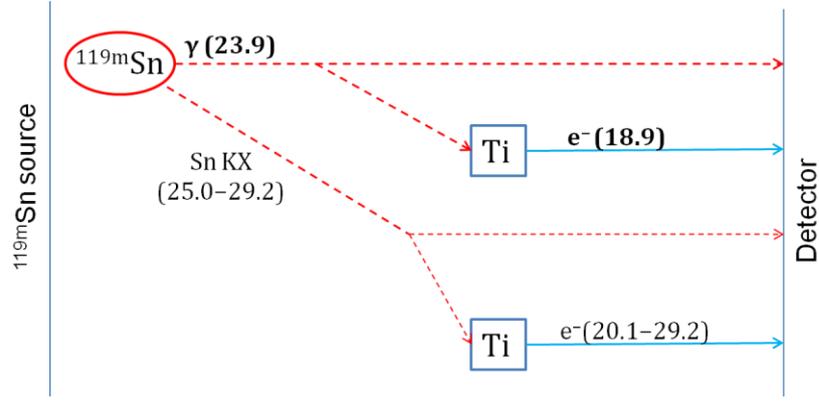

**Fig. 3.** Photons and electrons of energy around 18.9 keV emitted from the $^{119m}$Sn/Ti photoelectron source. Not shown, but included into calculations are interactions with $SnO_2$, Be exit window, C backing of the Ti convertor, and hydrocarbon contamination overlayer.

**3.2 Rate of zero-energy-loss photoelectrons**

The rate of zero-energy-loss photoelectrons of energy close to 18.6 keV recorded in the KATRIN monitor spectrometer can be approximately expressed in the form

$$N_{phe} = N_{dec}\ b_\gamma\ \Omega_c\ a_s\ \sigma_{ph}\ d_c\ s_{sh}\ \Omega_{sp}\ b_{zel}\ a_{cont}\ \eta_{det}\ , \qquad (2)$$

where  $N_{dec}$    is the decay rate of the radioactive source
         $b_\gamma$    is the branching ratio of γ- or X-rays
         $\Omega_c$    is a part of full solid angle seen by the convertor
         $a_s$    is the correction for photon absorption in the source
         $\sigma_{ph}$    is the photoelectric cross section in the convertor
         $d_c$    is the surface density of atoms in the convertor
         $s_{sh}$    is the correction for shake up/off processes in the photoelectron emission
         $\Omega_{sp}$    is the effective part of full solid angle seen by the spectrometer that takes into account angular distribution of photoelectrons emerging from the convertor
         $b_{zel}$    is a part of photoelectrons that did not suffer any energy loss while traveling through the convertor
         $a_{cont}$    is correction for intensity of zero-energy-loss photoelectrons passing through a hydrocarbon contamination layer on the convertor surface
         $\eta_{det}$    is the detector efficiency for photoelectrons of a particular energy.

The solid angle $\Omega_c = 0.16$ of 4π, obtained by simple Monte Carlo calculations for a disc geometry, comes from the capsule construction (see Fig. 1). The corrections for photon absorption in the sources $a_s$, were based on the assumption of a homogenous distribution of $^{241}$Am atoms within 1mm thick layer of Si (simulating absorption in enamel) and homogeneous distribution of $^{119m}$Sn in $SnO_2$. Photon attenuation through 1 mm thick beryllium exit window was considered, too. The spread of photon directions was determined by construction of capsules. For useful γ-rays of 26.3 keV populated in $^{241}$Am decay, the corrections for absorption in americium, silicon and beryllium were 0.47, 0.76 and 0.996, respectively, yielding $a_s$ = 0.36. The correction for the absorption of 24 keV γ-rays in SnO$_2$ and Be amounted to 0.67 and 0.994, respectively. Thus $a_s$ = 0.67 for the $^{119m}$Sn source.

**Table 1.** $^{241}$Am and $^{119m}$Sn sources of γ- and X-rays

| Photon emitter | $^{241}$Am | $^{119m}$Sn |
|---|---|---|
| ***Decay properties:*** | | |
| Half-life | 432.6 y | 293.1 d |
| Energy of useful γ-rays, $E_\gamma$ (keV) | 26.345 | 23.870 |
| Branching ratio of useful γ-rays, $b_\gamma$ | 0.024 | 0.165 |
| Energy of disturbing X-rays (keV) | 20.8–21.5 | 25.0–29.2 |
| Branching ratio of disturbing X-rays, $b_X$ | 0.358 | 0.276 |
| ***Commercial γ-ray source:*** | | |
| Source activity, $N_{dec}$ (GBq) | 1.1 | 0.55 |
| Diameter of active part (mm) | 7.2 | 10 |
| Thickness of Be exit window (mm) | 1 | 1 |
| Producer | Nuclitec, GmbH | Ritverc, GmbH |
| ***Assumed source properties:*** | | |
| Specific activity GBq g$^{-1}$ | 123 | 11 |
| Average thickness of active part | 18 μm of $^{241}$Am + 1mm of Si | 91 μm of SnO$_2$ |
| Mean free path of useful γ-rays, $E_\gamma$ | 13.8 μm in Am; 2.5 mm in Si; 0.33 m in Be | 148 μm in SnO$_2$; 0.24 m in Be |
| Correction for absorption of useful γ-rays within the source, $a_s$ | 0.36 | 0.67 |
| Rate of useful γ-rays of energy $E_\gamma$ seen by the convertor (s$^{-1}$) | 1.4·10$^6$ | 9.5·10$^6$ |

The K-shell photoelectric cross sections $\sigma_{ph}(K)$ = 1.09·10$^{-21}$ cm$^2$ and 6.55·10$^{-22}$ cm$^2$ for 26.345 keV and 23.870 keV γ-rays impinging on cobalt and titanium convertors, respectively, were determined from tabulations [36]. The 85% of zero-energy-loss photoelectrons originate in the convertor surface layer of the thickness equal to 2$\lambda_{inel}$, the double of the electron mean free path for inelastic scattering. Thicker convertors would predominantly increase background under the monitoring lines at 18.6 and 18.9 keV, respectively, due to inelastically scattered photoelectrons of higher energies (see section 3.3). Therefore, we assumed in our calculations the thicknesses of 31 and 54 nm (the relevant surface atomic densities $d_c$ = 2.7·10$^{17}$ atoms cm$^{-2}$ and 3.0·10$^{17}$ atoms cm$^{-2}$) for cobalt and titanium convertors, respectively. Such convertors would not be self supporting; e.g. the thinnest Co foil produced by Goodfellow [37] is 3 μm thick, while 100 nm cobalt layer is supported by a 3.5 μm Mylar foil. Since such a foil might be damaged by γ- and X-ray radiation during a long time operation, it would be better to deposit Co and Ti layers of requested thickness on 1 μm thick carbon foil produced by Acf-metals [38]. Such a foil would decrease the flux of useful γ-rays irradiating convertor in a negligible way but the photoelectrons ejected by X-rays from the carbon K-shell could increase the background.

Shake up and shake off processes, in which the energy available for one photoelectron is transferred to two bound electrons of the convertor atom, cause further decrease of intensity of zero-energy loss photoelectrons. We have estimated magnitude of the effect, $s_{sh}$ = 0.7 for Co and $s_{sh}$ = 0.64 for Ti, using the shake-off probabilities after $β^-$ decay of free atoms [39]. The shake up/off probability in metals is probably larger. Recently, the shake-off of conductivity electrons in metals caused by nuclear decay was theoretically examined [40] but relevant probabilities are not yet available. The effect was investigated experimentally e.g. in the decays of $^{57}$Co → $^{57}$Fe [41] and $^{83m}$Kr → $^{83}$Kr [17]. Our calculation of the effective spectrometer angle $Ω_{sp}$ refers to present adjustment of the Mainz MAC-E-filter with entrance polar angle of ±35°. In order to incorporate angular distribution of photoelectrons ejected from the convertor by a wide photon beam, we utilized tabulation [42]. Since the values presented there for photon energies of 100–5000 eV exhibit small energy dependence, we assumed that the parameters extrapolated to energies around 20 keV are applicable also for our estimates of expected photoelectron angular distribution.

A part of the 18.6 keV (18.9 keV) photoelectrons produced in the volume of the cobalt (titanium) convertor, and emitted into the spectrometer entrance angle without any extrinsic energy loss, was estimated to be $b_{zel}$ = 0.42 (0.41). Further, we assumed the presence of a hydrocarbon contamination layer on the convertor surface and approximated it by the 3 nm thick carbon layer. The correction factors $a_{cont}$ were determined using the electron inelastic mean-free-paths in Table 2. Finally, we have assumed the efficiency of the Si detector of the MAC-E filter for 18.6 keV electrons to be $η_{det}$ = 0.8.

Using the Eq. 2 and parameters in Tables 1 and 2 we predict the detected rate of monoenergetic photoelectrons of 18.632 keV energy, produced by the 1.1 GBq $^{241}$Am/Co source, to be roughly 9 s$^{-1}$. Similarly, we predict the detected rate of monoenergetic 18.901 keV photoelectrons from the 0.55 GBq $^{119m}$Sn/Ti source to be roughly 31 s$^{-1}$. Regardless lower source activity $N_{dec}$ and lower photoelectric cross section $σ_{ph}$, the latter source provides higher photoelectron rate $N_{phe}$ due to higher branching ratio $b_γ$ of useful γ-rays and their lower self absorption $a_s$ within the source material. In order to get rid of background caused by the photoeffect of Am L X-rays on cobalt convertor, we examined also the $^{241}$Am/Ti source (see section 3.3).

**3.3. Background rate of photoelectrons and of direct γ- and X-rays**

The $^{241}$Am → $^{237}$Np and $^{119m}$Sn → $^{119}$Sn decays are accompanied by emission of various γ- and X-rays [30] that produce disturbing photoelectrons in cobalt and titanium convertors, respectively. Since the MAC-E filter is an integrating electron spectrometer, the electrons of energy slightly higher than that of the monitoring ones will cause disturbing background. Some of these events can be eliminated by discriminating the amplitude of the detector signal. The electrons of much higher energy will not be integrated. Namely, they will not fulfill the condition for adiabatic motion in guiding magnetic field of the MAC-E-filter and will not reach the detector.

In our estimate of the photoelectron background rate, we first considered four lines of the neptunium L X-rays with energies 20.784–21.491 keV ejecting photoelectrons from all atomic subshells of cobalt except the K-shell. There is a relatively strong self absorption of photons in commercial $^{241}$Am source (see section 3.1.) that can be accompanied by emission of americium L X-rays. Their photoelectric absorption on the L and higher atomic subshells of cobalt may be another source of background. We considered six lines of Am LX-rays with energies from 18.856 to 22.836 keV producing disturbing photoelectrons with energies of 18.751–22.832 keV. In order to determine the Am L X-ray intensities the photoelectric cross sections for relevant atomic subshells of americium were estimated from numerical values of the total cross sections and graphs in the data base [43]. We assumed that the $\sigma_{subshell}/\sigma_{total}$ ratios at the energy thresholds are applicable also for higher energies.

**Table 2.** Sources of monoenergetic photoelectrons

| γ-ray emitter/Convertor | $^{241}$Am/Co | $^{119m}$Sn/Ti | $^{241}$Am/Ti |
|---|---|---|---|
| ***Photoelectron convertor:*** | | | |
| K-shell electron binding energy, $E_{b,F}$ (keV) | 7.709 | 4.966 | 4.966 |
| Natural width of K-shell vacancy (eV) | 1.3 | 0.94 | 0.94 |
| Photoelectron energy, $E_{kin}$ (keV) [a] | 18.632 | 18.901 | 21.376 |
| Electron inelastic mean-free-path in the convertor, $\lambda_{inel}$ (nm) | 15.5 | 27 | 24 |
| Convertor thickness (nm) | 31 | 54 | 48 |
| Electron inelastic mean-free-path in carbonaceous contamination (nm) | 32 | 39 | 36 |
| ***Photoelectric process:*** | | | |
| γ-ray solid angle seen by convertor, $\Omega_c$ | 0.16 | 0.16 | 0.16 |
| Photoelectric cross section for photons of useful energy, $\sigma_{ph}$ (K) ($10^{-24}$cm$^2$) | 1090 | 655 | 490 |
| Surface density of atoms in the convertor of thickness, $d_c = 2\lambda_{inel}$ ($10^{17}$ cm$^{-2}$) | 2.7 | 3.0 | 2.7 |
| Correction for shake up/off process in the convertor, $s_{sh}$ | 0.70 | 0.64 | 0.64 |
| Part of zero-energy-loss electrons, $b_{zel}$ that left convertor | 0.42 | 0.41 | 0.41 |
| Correction for electron intensity loss in contamination overlayer, $a_{cont}$ | 0.90 | 0.92 | 0.88 |
| ***Mainz electron spectrometer:*** | | | |
| Spectrometer work function, $\varphi_{sp}$ (eV) | 4 | 4 | 4 |
| Spectrometer solid angle $\Omega_{sp}$ including photoelectron angular distribution (% of 4π) | 9 | 9 | 9 |
| Electron detector efficiency, $\eta_{det}$ | 0.80 | 0.80 | 0.80 |
| Expected photoelectron rate, $N_{phe}$ (s$^{-1}$) | 8.7 | 31 | 3.4 |

[a] assuming $\varphi_{sp}$ = 4 eV and $C$ = 0 in Eq. 1.

In the case of the $^{119m}$Sn/Ti source, we similarly considered six lines of Sn KX-rays with energies of 25.044 – 29.176 keV ejecting photoelectrons with energies of 20.074 – 29.172 keV from various atomic subshells of titanium.

Resulting predictions of the rates of zero-energy-loss photoelectrons intended for checking the KATRIN energy scale as well as those producing disturbing background in detector of the MAC-E-filter are depicted in Figs. 4 and 5 for the $^{241}$Am/Co and $^{119m}$Sn/Ti sources, respectively.

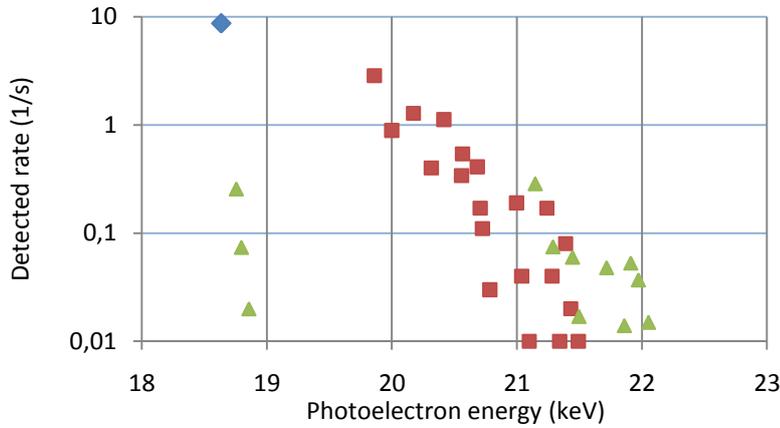

**Fig. 4.** Prediction of the detected photoelectron rates of the $^{241}$Am/Co source for the 31 nm thick cobalt convertor and MAC-E filter adjusted to 18.6 keV. The rate of the 18.632 keV zero-energy-loss photoelectrons (rhomb) is shown together with the rate of background photoelectrons created by Np L X-rays (squares) and Am L X-rays (triangles).

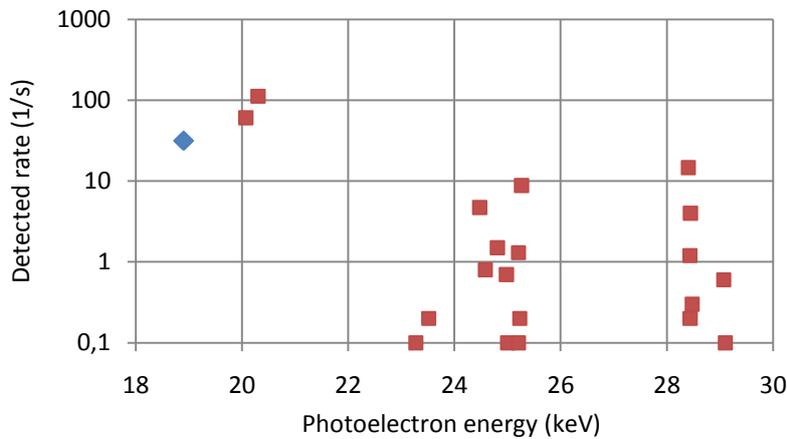

**Fig. 5.** Prediction of the detected photoelectron rates of the $^{119m}$Sn/Ti source for the 54 nm thick titanium convertor and MAC-E filter adjusted to 18.9 keV. The rate of the 18.901 keV zero-energy-loss photoelectrons (rhomb) is shown together with the rate of background photoelectrons created by Sn K X-rays (squares).

In the case of $^{241}$Am/Ti source, the background induced by Am LX-rays is suppressed. However, the price for lower background would be a lower yield of zero-energy-loss photoelectrons since the cross section $\sigma_{ph}$(Ti) for the 26.3 keV photons is 2.2 times smaller than $\sigma_{ph}$(Co). In addition, photoelectrons ejected by the 26.345 keV γ-rays of $^{241}$Am from the titanium K-shell would have energy of 21.376 keV, a bit too far from the endpoint energy of the tritium β-spectrum of 18.575 keV.

Finally, we estimated background caused by photoelectrons emitted from a carbon foil supporting extremely thin cobalt and titanium convertors. Main contribution comes from the photoeffect of neptunium and americium L X-rays and tin K X-rays on the carbon K-shell. Regardless of low atomic number of carbon, the cross section $\sigma_{ph}$ amounts 4 – 2 barn/atom for 21–25 keV photons. Thus the disturbing photoelectrons produced in 1 μm carbon backing cannot be neglected. (For aluminum backing the effect would be even worse since corresponding $\sigma_{ph}$ is about 30 times larger). The energy losses of photoelectrons within the carbon backing were treated by means of the stopping power reaching 2.5 keV/μm. We have verified the applicability of this approach by several detailed Monte Carlo simulation of electron scattering in the backing. The MC calculations of low energy tails of the photoelectron lines produced by 59.6 keV γ-rays of $^{241}$Am in the carbon backing showed that their contribution to background spectrum in the region of monitoring photoelectrons is negligible. Photoelectrons ejected from Ti convertor by 65.7 keV γ-rays of $^{119}$Sn can also be neglected since the relative intensity of these photons is only 2·10$^{-4}$ per $^{119m}$Sn decay [32].

A part of background is caused by direct hitting of γ- and X-rays on the detector regardless of its 4.4 m distance from the $^{241}$Am or $^{119m}$Sn sources (solid angle of 2.6·10$^{-7}$ of 4π).

Discriminator levels applied to semiconductor detector allow further reduction of background. We have considered two windowless Si detectors of 80 mm$^2$ area:
– 0.5 mm thick windowless PIN photodiode S359.60-06 with FWHM = 3 keV for 20 keV electrons, made by Hamamatsu [44]
– 5 mm thick Si(Li) detector with assumed FWHM = 0.8 keV for 20 keV electrons[2].

The background rates were calculated also according to Eq. 2 but the corrections $s_{sh}$, $b_{zel}$ and $a_{cont}$ were omitted since the electrons that suffered energy loss contribute to the background in the MAC-E filter, too.

The reduced background rates, corresponding to the spectrum regions of <$E_{kin}$ – 3 keV, $E_{kin}$ +3 keV> and <$E_{kin}$ –0.8 keV, $E_{kin}$ +0.8 keV> around energies $E_{kin}$ of monitoring photoelectron lines (also in keV), are summarized in Table 3. The rates were calculated assuming simplified triangle shape of the detector response functions and include all spectrum components exhibited in Figs. 4 and 5.

## 3.4. Monte Carlo simulation of photoelectron spectra

Simplified calculations described in section 3.2 enabled us to clarify the role of individual processes listed in Eq. 2 that determine the expected photoelectron rate, $N_{phe}$. This quantity as well as the spectrum shape can be predicted by means of the Monte Carlo method. We applied the MC code [46] which treats individual elastic and inelastic scatterings of electrons in solids and can provide the energy and angular distribution of photoelectrons leaving the convertor. The code proved useful in the low-energy region; e.g. the experimental β- spectrum of $^{241}$Pu was described down to 2 keV without any artificial correction [47].

Photoelectrons created in a convertor experience both elastic and inelastic scattering during their exit from the source. The process is characterized by elastic and inelastic electron mean free path, $\lambda_{el}$ and $\lambda_{inel}$, respectively. Reliable values of $\lambda_{el}$ can be deduced from calculations of the cross sections for elastic scattering on particular atoms [48]. Experimental and theoretical values of $\lambda_{inel}$ are summarized in a number of papers, e.g. [49–51], predominantly in the energy region up to 2 or 10 keV. In the case of cobalt, $\lambda_{inel}$ = 15.5 nm at 18.6 keV was deduced as the mean of values for Ni and Cu in [52].

---

[2] This energy resolution was estimated from our experience with the 3 mm thick Si(Li) detector of 20 mm$^2$ area, developed formerly by Miloš Vidra at the Nuclear Research Institute at Řež. Using $^{57}$Co vacuum evaporated source, we measured FWHM = 0.59 keV for 13.6 keV conversion electrons.

**Table 3.** Expected background rate (s$^{-1}$) of the photoelectron detector in energy region of the monitoring line

| Photoelectron source | $^{241}$Am/Co | | $^{119m}$Sn/Ti | | $^{241}$Am/Ti | |
|---|---|---|---|---|---|---|
| MAC-E-filter set to (keV) | 18.6 | | 18.9 | | 21.4 | |
| Detector FWHM (keV) | 0.8 | 3.0 | 0.8 | 3.0 | 0.8 | 3.0 |
| Photoelectrons: | | | | | | |
| Np L-series | 1.0 | 5.6 | | | 0.0 | 0.0 |
| Am L-series | 0.2 | 0.5 | | | 0.2 | 0.2 |
| Sn K-series | | | 12 | 160 | | |
| low-energy tails | 0.0 | 0.1 | 0.0 | 0.0 | 0.0 | 0.0 |
| convertor backing | 1.8 | 3.6 | 0.0 | 0.1 | 0.1 | 0.3 |
| Photons: | | | | | | |
| γ-rays | 0.0 | 0.0 | 0.0 | 0.2 | 0.0 | 3.5 |
| Np LX-rays | 3.8 | 6.1 | | | 2.3 | 3.0 |
| Am LX-rays | 0.2 | 0.2 | | | 0.1 | 0.1 |
| Sn KX-rays | | | 0.0 | 2.0 | | |
| Effect / background | 1.2 | 0.5 | 2.6 | 0.2 | 1.3 | 0.5 |

In our MC simulations, physical assumptions and geometrical arrangement of the $^{241}$Am/Co photoelectron source were the same as in section 3.2. At the beginning of simulated random history, the 26.3 keV photon is generated at any point within the volume of 1 mm thick enamel disc containing $^{241}$Am atoms. Naturally, isotropic angular distribution of emitted photons was supposed. Possible photon absorption within enamel (of Si nature) and Be window as well as self absorption by Am atoms were taken into account. Only the photons which can exit from the monel source capsule and irradiate the Co convertor were followed. Angular distribution of photoelectrons ejected from the Co K-shell was included, too [42]. The path of the photoelectron within the Co convertor as well as in 3 nm thick contamination overlayer on cobalt surface was simulated by means of MC code [46]. This code treats individual elastic and inelastic electron collisions and provides energy and angular distribution of photoelectrons leaving the convertor.

The role of electron energy losses within cobalt convertor of the $^{241}$Am/Co source is illustrated in Fig.6 for thin and thick convertor. Spectra of K-26 photoelectrons emitted into the entrance angle of 9 % of 4π around normal to the Co surface (present adjustment of the Mainz spectrometer) are displayed. In these MC simulations, the shake up/off processes were taken into account with the 30 % probability ($c_{sh}$ = 0.7 in Eq. 2 and Table 2) and corresponding energy distribution according to approach described in [46]. The intensity of the zero-energy-loss peak at 18.632 keV in both spectra is about the same, as expected. Detailed calculations revealed that a scatter of the points above 18.55 keV expresses the structure of the energy loss function of cobalt. This function described the probability distribution of electron energy losses in the case of inelastic scattering.

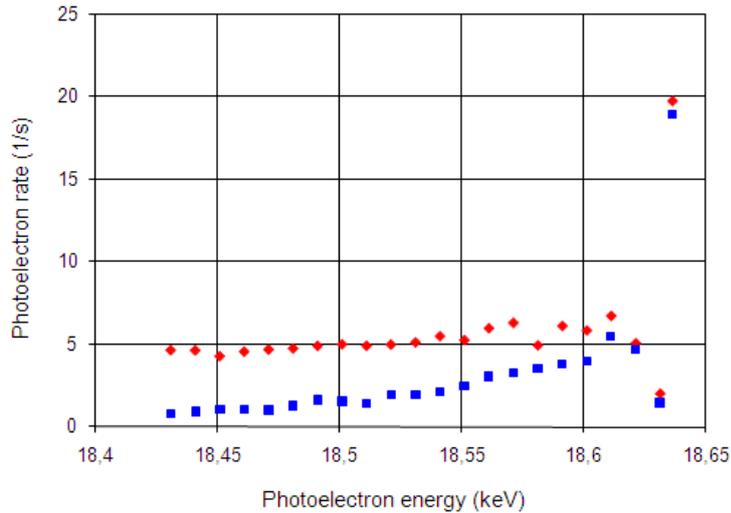

**Fig. 6.** Monte Carlo simulated spectra of the K-26 photoelectrons emitted from the $^{241}$Am/Co source for 31 nm (squares) and 5 000 nm (rhombs) convertors. In order to get the spectrum for the thick convertor the history of $2 \cdot 10^9$ γ-quanta of 26.345 keV energy, born inside the $^{241}$Am commercial source, had to be followed.

Comparison of simulated photoelectron spectra with those measured in this work is depicted in Figs. 8– 11.

## 4. Experimental
### 4.1. γ-ray spectroscopy of $^{241}$Am source

Calculations in section 3.2 indicate substantial absorption of 26.3 keV photons within the 1.1 GBq $^{241}$Am source [33]. In order to check the correction factor $a_s$ in Eq. 2 we compared γ-ray spectrum of this sealed source with that of the $^{241}$Am reference source Eg3-454-36 (product of the Czech metrology institute [53]) exhibiting very small photon absorption. In fact, the 1.2 mm thick source envelope made of polymethyl methacrylate decreases intensity of 26.3 and 59.6 keV photons by factors of 0.981 and 0.999, respectively.

Measurements were carried out using the Ortec apparatus equipped with the 24% HPGe detector in a horizontal cryostat. The distance between the source and the detector was set to 320 cm. From comparison of the two γ-spectra it follows that absorption in americium, enamel and beryllium of the sealed $^{241}$Am source reduces the flux of 26.3 keV photons emitted in direction of the source axis by a factor of 2.09 ± 0.10 more than the flux of 59.6 keV photons. Agreement with the calculated value of 2.0 demonstrates that our assumptions about the internal structure of the sealed $^{241}$Am source, as listed in Table 1, are realistic.

The activity of the $^{241}$Am standard (referred to 1.3.2009) was 675 ± 12 kBq. Using this value and the ratio of 59.6 keV γ-line intensities of the two sources corrected for absorption effects, we determined activity of the $^{241}$Am sealed source to be 1.23 ± 0.04 GBq. This is in accord with the producer value of 1.11 ± 0.11 GBq.

## 4.2. Preliminary measurements of the $^{241}$Am/Co source with the Řež electron spectrometer

Three convertor thicknesses available from Goodfellow [37] were examined: 5, 3 and 0.1 μm. Drs. J. Zemek and K. Jurek from the Physics Institute of the ASCR in Prague investigated the 5 μm thick Co foil by means of photoelectron spectroscopy (XPS) and a scanning electron microscopy. The photography of the more shiny side of the foil taken with the magnification of 4 000 is exhibited in Fig. 7. The XPS examination revealed that the foil as received was covered by an oxide layer thicker than 6 nm since all cobalt was bounded to oxygen while there was no signal from the Co-Co bonds. Using an electron microprobe and assuming tabulated density and stoichiometry of CoO, the thickness of the oxide layer was determined to be 24 nm.

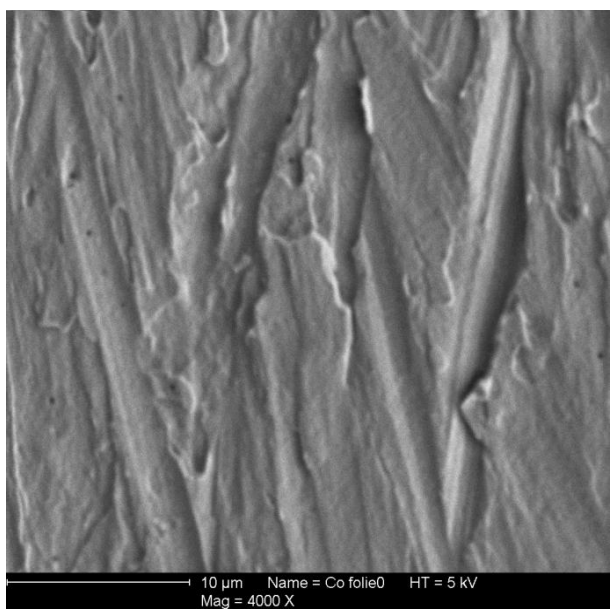

**Fig. 7**. Surface of the shiny side of the 5μ thick cobalt foil as received from the producer [37] seen in an electron microscope with magnification of 4000.

Preliminary photoelectron measurements were carried out with the ESA 12 spectrometer [54, 55]. This is a differential double-pass cylindrical mirror analyzer operating in its basic mode with the transmission of 0.74 % of 4π and the energy resolution $\Delta E/E$ = 0.011. Instrumental resolution better than 1 eV was reached with the pre-retardation system (see e.g. [23]) but this mode is not accessible to 18.6 keV electrons. Only electrons emitted in a narrow angular interval around 40.8˚ with respect to the source surface normal are analyzed. A windowless channel electron multiplier serves as the electron detector. Spectrometer operates at vacuum of $5 \cdot 10^{-8}$ mbar.

The $^{241}$Am source [33] of 1.1 GBq activity was mounted into the spectrometer holder together with a cobalt convertor in a closed disc geometry. Due to a low transmission of the instrument only photoelectrons of 4.5 – 14 keV energy created by L X-rays of daughter $^{237}$Np were investigated. Examples of measured spectra are shown in Fig. 8 and 9 together with MC predictions described in section 3.3. The only free parameter in depicted simulated spectra is the absolute efficiency of the electron detector $\eta_{det}$ (about 10% at 10 keV, in a reasonable agreement with previous determinations). Energy dependence of $\eta_{det}$ was deduced from previous determinations [54, 57].

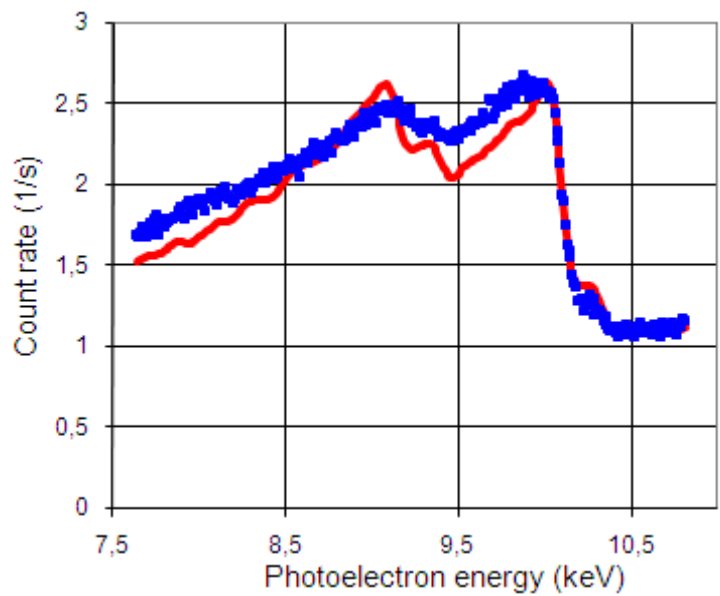

**Fig. 8.** A part of the $^{241}$Am/Co photoelectron spectrum taken with the 5µm thick convertor in the β-ray spectrometer of the NPI at Řež. The instrumental resolution was set to 110 eV (FWHM) at 10 keV and the transmission was 0.74 % of 4π. The solid line represents a Monte Carlo simulation of photoelectrons ejected from the cobalt convertor by $^{237}$Np L X-rays populated in the $^{241}$Am α-decay (see section 3.3). The detector background of 0.4 s$^{-1}$ was subtracted.

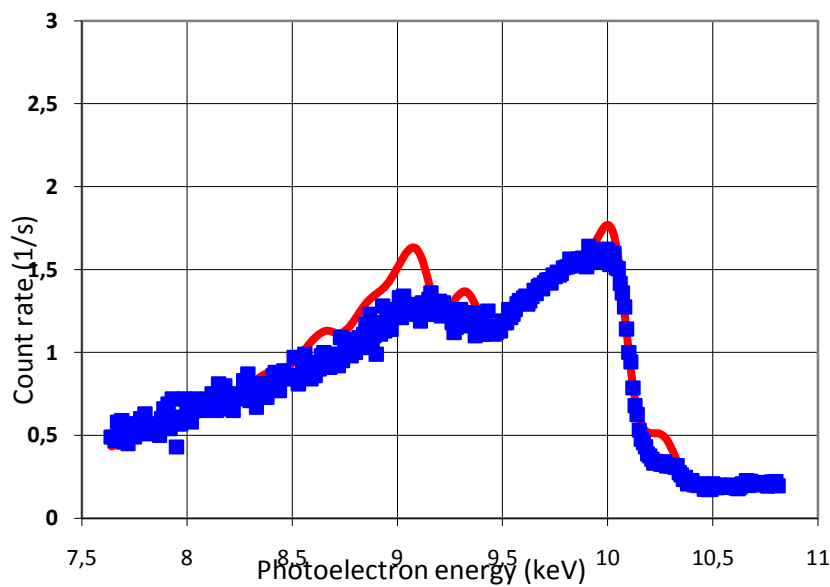

**Fig. 9.** The same photoelectron spectrum as in Fig. 8 taken with the 0.1 µm cobalt convertor. Note substantially lower background than in the case of 5 µm convertor.

There is a reasonable agreement between measured and simulated spectra. Comparing measured spectra for 5 µ and 0.1 µ Co convertors in Fig. 8 and 9 we see that in both cases the peak at 10 keV has intensity of about 1.5 s$^{-1}$. As expected, the background is much smaller for thinner convertor. Details of the measurements are described in [56].

## 4.3. Pilot electron measurements at high resolution with the Mainz spectrometer

The Mainz spectrometer enabled us to examine the 18.632 keV zero-energy-loss photoelectrons of the $^{241}$Am/Co source with the instrumental resolution of 1 eV. Thanks to magnetic adiabatic collimation this electrostatic filter exhibits high transmission which is of a great importance in studies of low-intensity features. On the other side, the integral character of the recorded spectra means that low-energy lines may be disturbed by an increased background caused by the electrons of higher energy. This drawback may be partly compensated by a proper choice of discrimination levels applied to signals of the electron detector.

The $^{241}$Am/Co source with 3μm thick cobalt foil was first inserted into the spectrometer in the direct geometry, i.e. with its axis identical with the spectrometer axis. A small effect of about 20 s$^{-1}$ superimposed on high background of about 3200 s$^{-1}$ was observed in the region of 18.6 keV. Only a few percent of the background can be caused by direct γ- and X-rays. When the source was tilted by 65° with respect to the spectrometer axis, the background decreased to about 310 s$^{-1}$ as seen in upper part of Fig. 10.

Monte Carlo simulations revealed that a part of the background is caused by the K-59.5 photoelectrons (ejected from the cobalt K-shell by 59.6 keV γ-rays of $^{241}$Am with original kinetic energy of 51.8 keV) that suffered a large energy loss within relatively thick Co convertor. In the case of the direct geometry, the contribution of the 59.5 photoelectrons to background is roughly 1000 s$^{-1}$ in the 3 keV interval above 18.6 keV. In the tilted geometry this contribution is roughly 200 s$^{-1}$. This is a further argument in favor of thinnest possible convertor as considered in section 3.2.

A part of the background is probably caused by secondary electrons knocked from the inner surface of the spectrometer vacuum vessel by γ- and X-rays of 1.1 GBq $^{241}$Am source. These electrons born around the analyzing plane of the MAC-E filter may be accelerated and reach the detector with kinetic energy about 18.6 keV. The effect can be removed by wire electrodes put on a slightly more negative potential than that on the vessel as intended for the main KATRIN spectrometer [12].

We attempted to increase intensity of the K-26 photoelectrons by cleaning the cobalt foil in situ by means of Ar ion etching. This procedure would in addition provide a well defined and reproducible convertor surface. The Ar ions with the energy of 100 eV and current density of 100 μA cm$^{-2}$ fell on the Co foil with the incident angle of 45°. The thickness of sputtered layer was estimated to be 16 nm in the case of cobalt oxides. The photoelectron spectra prior and after the convertor sputtering are compared in Fig. 10. One can see that the Ar etching improved the sharpness of the effect and decreased background by a few percent.

A substantial background reduction was reached by tuning adiabacity of the MAC-E filter spectrometer changing the various magnetic field ratios. In this way some part of electrons with higher kinetic energy did not follow the guiding magnetic field lines and they did not hit the detector. After some iterative procedure optimal setting was accomplished. Resulting spectrum of the K-26 photoelectron line of the $^{241}$Am/Co source is shown in Fig. 11. The zero-energy loss peak of about 2 s$^{-1}$ intensity is superimposed on background of about 16 s$^{-1}$ intensity. Thus the background was reduced by a factor of 20 while the effect decreased about three times.

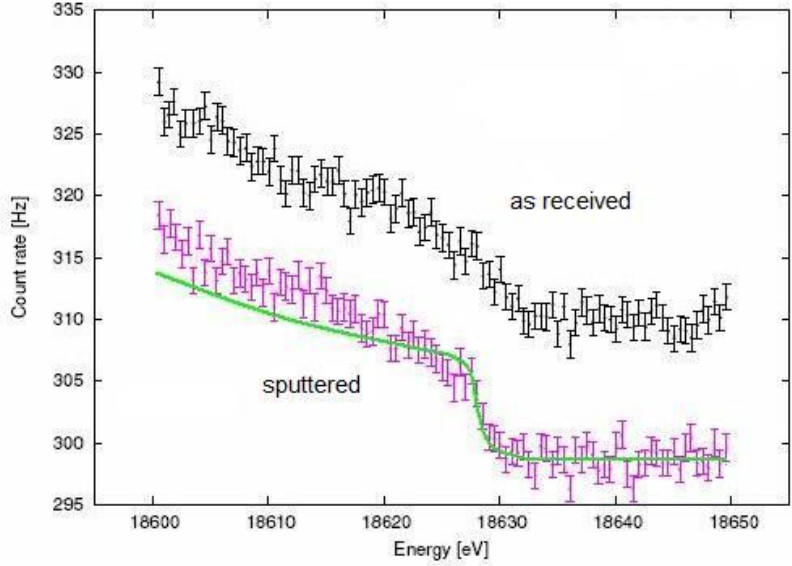

**Fig. 10.** The K-26 zero-energy-loss photoelectrons of the tilted $^{241}$Am/Co source measured with the Mainz spectrometer set to instrumental resolution of 1 eV (full width). The upper spectrum corresponds to 3 μm thick cobalt convertor as received; the lower spectrum was recorded after etching the convertor surface with argon ions (see text). Solid line shows the result of the Monte Carlo simulation for the source and spectrometer parameters given in tables 1 and 2. In particular, the solid angle $\Omega_{sp}$ was equal to 9% of 4π. A constant background was the only free parameter of the fit.

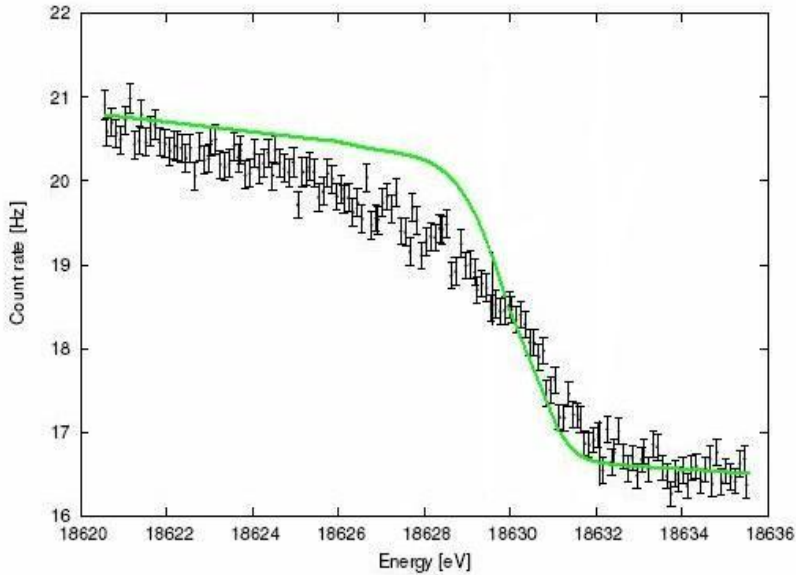

**Fig. 11.** The K-26 zero-energy-loss photoelectrons recorded by means of the Mainz β-ray spectrometer with tuned adiabacity and 1 eV instrumental resolution. The $^{241}$Am/Co source was tilted by 65° with respect to the spectrometer axis and surface of the 3 μm thick cobalt convertor was etched by argon ions. Solid line shows the result of Monte Carlo simulations for the same parameters as in Fig. 10, except the solid angle $\Omega_{sp}$ that was assumed to be 4.5% of 4π for tuned adiabacity of the MAC-E-fiter. There was no normalization fitted to the data and still agreement to the data within a factor of 2 is observed. This is very reasonable considering the significant

systematic uncertainties of our accepted solid angle in the "adiabaticity-tuned" mode as well as the Monte Carlo uncertainties for the tilted source.

## 5. Discussion and conclusions

The sources of monoenergetic electrons based on the photoelectric effect of nuclear γ-rays fulfill requirements for appropriate energy and small natural width. However, the sources provide rather low photoelectron rate shown in table 2. In the case of $^{241}$Am, it is due to low branching ratio of 26.3 keV photons, large half-life and pronounced self absorption of low energy photons in americium (Z = 95). Photoelectron rate of the $^{119m}$Sn/Ti source is limited mainly by the specific activity of $^{119m}$Sn achievable in a nuclear reactor.

A clear advantage of the investigated photoelectron sources in their potential stability and definition in energy. The combination of sharp and precisely known γ-ray energy, together with a reliable electron binding energy (corresponding to the well defined environment of an ion-bombardment cleaned metallic convertor) should lead to reliable electron kinetic energies based on a nuclear and atomic standard. Combining such a photoelectron source with electrostatic retardation spectrometer (e.g. a MAC-E-filter) may even help in development of an absolute high voltage standard. Such standard would be beneficial for KATRIN, too.

Common drawback of the radionuclide sources is that they emit also γ- and X-rays causing disturbing background in interested energy region of the photoelectron spectrum. A part of background originates in the photoeffect in convertors and their backings; the second part is due to the sensitivity of semiconductor detectors to a direct photon beam. In order to suppress the first contribution to background we propose to apply very thin convertors saving 85% of useful zero-energy-loss photoelectrons (with regard to an infinitely thick convertor). At the same time, the thin convertor minimizes rate of inelastically scattered photoelectrons (see Table 3). We also chose the thinnest self supporting low-Z backing.

The second source of background could, in principle, be eliminated by appropriate shielding of X-rays directed to the detector while eliminating only a very small part of solid angle for photoelectrons guided by a magnetic field of the MAC-E filter. One could also evaporate a thin Co convertor directly on the Be exit window and thus avoid disturbing photoelectrons from the C backing. However, the $^{57}$Co source evaporated on the Be backing exhibited continuous deterioration of the 7 keV conversion electron line [58], probably due to diffusion of $^{57}$Co atoms into Be. In our case, such diffusion would decrease already low intensity of the 18.6 zero-energy-loss photoelectrons emitted from the $^{241}$Am/Co source.

The X-ray excited photoelectron spectra of the $^{241}$Am/Co source measured with a moderate resolution is in reasonable agreement with our MC prediction. The measured flux of 26.3 keV photons emitted from the commercial $^{241}$Am source corresponds to the expected value. These two facts indicate that we did not miss in our simplified calculations any important feature. The pilot high-resolution measurements of the $^{241}$Am/Co source, carried out before reconstruction of the Mainz MAC-E filter for the KATRIN monitoring spectrometer, demonstrated importance of appropriate thickness of a photoelectron convertor.

The discrepancy between the rate of the 18.6 keV zero-energy-loss photoelectrons of the $^{241}$Am/Co source obtained by Monte Carlo simulations (19 s$^{-1}$ in Fig. 6) and the estimate of 8.7 s$^{-1}$ (in Table 2) is not surprising. For a broad photoelectron beam, the simple model based on Eq. 2 cannot compete with detailed MC calculations. Still, the simplified calculations are useful since they clarify the role of individual processes within the source and convertor. On the basis of these calculations we conclude that

- the $^{119m}$Sn/Ti photoelectron source provides the highest rate of useful zero-energy-loss photoelectrons and the best effect/background ratio
- the $^{241}$Am/Ti source does not provide any substantial advantage in comparison with the two other photoelectron sources
- The effect/background ratio of registered zero-energy-loss photoelectrons can be substantially improved using the high-resolution Si(Li) detector in the MAC-E filter
- none of the considered photoelectron sources based on commercial sources of $^{241}$Am and $^{119m}$Sn provides counting rates of zero-energy-loss photoelectrons comparable with those of $^{83}$Rb/$^{83m}$Kr conversion electron sources.

The photoelectron sources differ principally from conversion electron ones. Contrary to open radioactive sources, one can expect that the γ-ray intensity of a sealed radiator is governed by its half-life only. The physical and chemical stability of a metallic convertor is much better than that of an extremely small amount of radioactive material in open conversion electron sources.

In order to determine position of the 17.8 keV monitoring line of the $^{83}$Rb/$^{83m}$Kr conversion electron source with requested uncertainty of σ = 10 meV one needs to scan its full zero-energy-loss profile typically for 1.5 hour. In the case of weaker photoelectron sources one could consider to measure only at three points of the monitoring line (top, inflexional point and background). Similar procedure is proposed e.g. for often checks of the column density of the windowless gaseous tritium source with an electron gun [12]. In the case of the $^{119m}$Sn/Ti source treated in section 3.2, the exposure time of 45 minutes at the inflexional point would be needed to recognize change of the counting rate corresponding to 10 meV line shift. Adding another 45 min for checking the top and background counting we come to the exposure time comparable with that of the $^{83}$Rb/$^{83m}$Kr source.

## Acknowledgement

We are obliged to Drs. J. Zemek and K. Jurek for investigation of a commercial cobalt foil by electron spectroscopy methods. Three of us (J. Kašpar, A. Kovalík and D. Vénos) are grateful to the Deutsche Forschungsgemainschaft for financial help that allowed them to participate in the measurements with the Mainz β-ray spectrometer. The Řež group was supported by the Academy of Sciences of the Czech Republic under contract ASCR IRP AV0Z10480505 and by the MŠMT grants LA318 and LC07050. The coauthors from the universities in Mainz and Münster acknowledge the support from the German Bundesministerium für Bildung and Forschung.